# Hyperacute pathophysiology of traumatic and vascular brain injury captured by ultrasound, photoacoustic, and magnetic resonance imaging

Ali Kamali[1*], Laurel Dieckhaus[1*], Emily C. Peters[2], Collin A. Preszler[1], Russel S. Witte[1,3,4], Paulo W. Pires[2**], Elizabeth B. Hutchinson[1**], Kaveh Laksari[1,5**]

1 – Department of Biomedical Engineering, University of Arizona College of Engineering, Tucson, AZ

2 – Department of Physiology, University of Arizona College of Medicine, Tucson, AZ

3 – Department of Medical Imaging, University of Arizona College of Medicine, Tucson, AZ

4 – College of Optical Sciences, University of Arizona, Tucson, AZ

5 – Department of Aerospace and Mechanical Engineering, University of Arizona College of Engineering, Tucson, AZ

*These individuals contributed equally as co-first authors.

**These individuals contributed equally as senior authors.

Highlight (50 characters): Imaging hyperacute pathophysiology of brain injury

# Abstract

Cerebrovascular dynamics and pathomechanisms that evolve in the minutes and hours following traumatic vascular injury in the brain remain largely unknown. We investigated the pathophysiology evolution within the first three hours after closed-head traumatic brain injury (TBI) and subarachnoid hemorrhage (SAH), two common traumatic vascular injuries, in mice. We took a multi-modal imaging approach using photoacoustic, color Doppler, and magnetic resonance imaging (MRI) in mice. Brain oxygenation (%$sO_2$) and velocity-weighted volume of blood flow (VVF) values significantly decreased from baseline to fifteen minutes after both TBI and SAH. TBI resulted in 19.2% and 41.0% ipsilateral %$sO_2$ and VVF reductions 15 minutes post injury while SAH resulted in 43.9% %$sO_2$ and 85.0% VVF reduction ipsilaterally ($p<0.001$). We found partial recovery of %$sO_2$ from 15 minutes to 3-hours after injury for TBI but not SAH. Hemorrhage, edema, reduced perfusion, and altered diffusivity were evident from MRI scans acquired 90-150 minutes after injury in both models although the spatial distribution was mostly focal for TBI and diffuse for SAH. The results reveal that the cerebral %$sO_2$ deficits immediately following injuries are reversible for TBI and irreversible for SAH. Our findings can inform future studies on mitigating these early responses to improve long-term recovery.

# Introduction

Vascular damage in the brain can occur over a range of etiologies, from trauma (1) to cerebrovascular disease (2,3), and the resulting rupture and hemorrhage of vessels is a major precipitating event that often leads to disability or death (4). It is known that the earliest outcomes following vascular injury are remarkably consequential, but the pathophysiology of this critical time window remains poorly understood. With a deeper knowledge about hemodynamic changes and cellular outcomes during the minutes and hours after vascular damage, diagnostic and therapeutic strategies may be developed to improve long-term outcomes.

Primary pathophysiology of traumatic injury to brain tissue stems from physical damage to the cerebral blood vessels, neurons, and glia, and is preventable but may not be treatable (5–7). Secondary pathophysiology begins immediately after the primary injury and spans from hyperacute (minutes to hours) to chronic (weeks to years) phases with different pathological profiles. Starting with the hyperacute period, hemorrhage (8,9), edema (10), and inflammation (11) emerge and evolve. Some cells are permanently lost, whereas others undergo reversible damage and may be amenable to recovery in later phases (12–14).

Subarachnoid hemorrhage (SAH) is caused by perforation or total rupture of the vascular wall, normally as a consequence of a punctured aneurysm or severe head trauma (15), leading to extravasation of blood into the subarachnoid space. Depending on the diameter of the ruptured vessel, size of fissure and clotting status of the individual, the amount of blood is enough to fill the subarachnoid space and sinuses, consequently blocking the normal flow of cerebral spinal fluid and drainage of the brain (16). Over time, the accumulated fluid leads to an increase in intracranial pressure, reducing perfusion pressure of the brain and culminating in secondary ischemia. Spontaneous contractions of cerebral vasculature, known as vasospasms, are also correlated with severe SAH and known to exacerbate the ischemic injury (17). There is limited information about the hyperacute period of secondary injury after TBI and SAH. However, the prominence of hemorrhage and other vascular injuries as well as their strong association with mortality (18,19) places these outcomes at the forefront of promising diagnostic targets in the minutes and hours following head trauma.

Vascular and tissue injury coexist with multiple alterations in physiological metrics in the brain that can be captured by a multi-modal imaging approach. Magnetic resonance imaging (MRI), in particular diffusion imaging, allows for differential and quantitative detection of cytotoxic and vasogenic edema, which are common features of brain injury. Restricted or enhanced water diffusion, normally represented by apparent diffusion coefficient (ADC) (20) or diffusion tensor Trace (21), can distinguish between these two types of edema. Time-of-flight MRI provides contrast between vessels containing flow of blood and surrounding tissues. Moreover, MR sequences such as arterial spin labeling (ASL) can reveal altered cerebral blood flow (CBF) following injury, which itself results in compromised oxygen delivery and consumption in tissues (22). Moreover, recent advances in photoacoustic (PA) imaging (23) and transcranial Doppler (24) enable the quantification of blood oxygen saturation ($\%sO_2$) and hemodynamics.

Although these imaging modalities could play a significant role in diagnosis and prediction of human outcomes, only sparse imaging data exists within the hyperacute phase after injury (25). Pre-clinical models, on the other hand, can provide a more comprehensive picture of the cerebrovascular dysfunction and metabolism alterations in this period. Rodent TBI studies show focal decreases in CBF within minutes to hours post-injury with recovery and increases seen on later days (26,27). This reduced CBF is accompanied by increased cerebral metabolism for up to six hours after injury followed by a metabolic depression period for up to 10 days (28). Recently, A mouse model of TBI showed a reduction in focal %$sO_2$ derived via photoacoustic imaging six hours following the injury, returning to normal the following day (23). A limited number of rodent TBI studies have characterized hyperacute cerebral edema and perfusion using MRI, starting from 1h post-injury and tracking the injury cascade over days (27,29). Overall, these studies reveal how a multimodal imaging approach can shed light on coexistence of cytotoxic or vasogenic edema with reduced cerebral CBF and %$sO_2$, indicating the interplay between vascular injury, ionic imbalances, and blood brain barrier disruption. However, none involved closed-head TBI models, which is the closest model to the vast majority of human non-piercing blunt head trauma (30). Moreover, to the best of our knowledge, no study has comprehensively explored cerebral hemorrhage, edema, perfusion, and oxygenation in the hyperacute stage following brain injury.

The objective of the current study was to identify the pathophysiology of two brain injury subtypes – SAH and moderate TBI – in the hyperacute period (minutes to hours) post injury in mice. We used a combination of ultrasound, color Doppler, PA, and MR imaging from baseline up to 3 hours post injury for this purpose. Implementation of SAH via an endovascular perforation model allowed us to isolate the early outcomes of severe vascular injury due to direct vascular rupture and evaluate those against vascular injury induced by a moderate closed-head controlled cortical impact TBI model. Taking a multi-modal imaging approach, we characterize the hyperacute changes in global and local injury markers including blood %sO$_2$, blood flow velocity, hemorrhage, edema, and CBF.

## Methods

### Animals

Male and female BALB/C mice (n = 24, 12 female, age on day of injury = 15.3 ± 2.6 weeks) were purchased from Jackson Labs and Charles Rivers Laboratories. The animals were divided into four groups: TBI, TBI-sham, SAH, and SAH-sham (three males and three females per group). All animal procedures were approved by the Institutional Animal Care and Use Committee (IACUC) at the University of Arizona and designed based on the Animal Research: Reporting of In Vivo Experiments (ARRIVE) guidelines.

### Injury and Scanning Timeline

For injury manipulations, and baseline and post-injury imaging sessions, each animal was deeply anesthetized by isoflurane inhalation. We used a 3% level of isoflurane for induction and kept the level in the 1-2.5% range with the goal of retaining the respiration rate in the 40-60 breaths per minute range. The carrier for the anesthetic was 100% oxygen for injury/sham procedures and 50% oxygen and 50% air for imaging delivered with a total flow rate of 1 L/min (31,32). We chose a combination of oxygen and air for the prolonged imaging sessions to minimize the compounding hyperoxic effects. We collected MRI baselines on a separate day prior to injury. On the day of injury, we removed the animal's scalp to minimize signal artifacts for obtaining photoacoustic scans. We collected ultrasound (US), color Doppler and photoacoustic (PA) scans (referred to US/PA in short) immediately prior to injury, then induced the injury (either TBI or SAH) and noted the time (the exact time of impact for TBI, or when SAH microfilament ruptured the vessel near the Circle of Willis). The sham protocols were similar to the injury protocols except for the head impact or vascular perforation. Approximately 15 minutes after the injury induction, we moved the animal back to the ultrasound instrument and collected the 15-minute post-injury US/PA scan. We transferred the animal to the MR scanner approximately 60 minutes post-injury and allowed the animal's vitals to stabilize. The MR scans began 90 minutes post-injury and took approximately 1 hour including setup and pre-scans. The animal transfer between the US and MRI scanners took less than one minute. After the MR scans ended, we transferred the animal back to the US/PA instrument to acquire the 3-hour post-injury US/PA scans (Figure 1A).

**TBI Model**

To induce moderate TBI, we delivered a previously established closed-head controlled cortical impact (chCCI) model in mice (33,34) (Figure 1B). We installed an impactor (Leica ImpactOne, Leica Biosystems, IL) onto a stereotactic frame. To ensure equivalent lateral impact to each mouse, we adjusted the impactor to an angle of 10 degrees tilted from the vertical direction. We then placed the animal inside the stereotactic frame and delivered anesthesia using a cone-shaped tube. To ensure head support without removing all degrees of freedom and elevate the skull to be level with the flat tip of the impactor, we positioned a 5-mm thick stack of folded paper towels underneath the mouse's head. We delivered the TBI injury 2 mm to the left of bregma using a 5-mm-diameter impactor tip, terminal velocity of 6 m/s, depth of 2 mm, and dwell time of 100 ms, based on previously published values (33,34). In the sham protocol, we kept the mice under anesthesia and placed them under the impactor for the same duration as the injury protocol but did not apply any impact. We verified location of impact and head movement trajectory by high-speed videography (Chronos 1.4, Kron Tech., Canada) at a temporal resolution of 10,000 frames/s.

## SAH Model

We performed subarachnoid hemorrhage on the mouse brain using a previously established endovascular perforation model (35) (Figure 1B). The surgery involved placing the mouse supine and making an initial midline incision along the neck. After incision, we dissected and permanently ligated the left common and external carotid arteries (CCA and ECA, respectively); and then closed the internal carotid artery (ICA) with a temporary suture. Following temporary closure of the ICA, we made a small incision along the CCA, between the CCA ligation point and the branching point of the ECA and ICA. We inserted a 5-0 nylon suture (Ethilon, Ethicon Inc., USA) with a non-coated blunt end through the CCA incision and advanced the filament through the reopened ICA into the brain until we observed resistance. After reaching a resistance point, we pushed the filament further into the brain which induced endovascular perforation of the Circle of Willis near the middle cerebral artery

(MCA) (35), at which point we retracted and removed the filament from the artery and permanently ligated the CCA posterior to the incision for filament insertion. We performed SAH-sham surgeries by dissecting the CCA and its branches, permanently ligating the CCA and ECA, and temporarily suturing the ICA, which we then removed at the end of surgery. We did not use intracranial pressure monitoring due to its invasive nature. Nonetheless, other studies have forgone the use of intracranial pressure monitoring during SAH surgery and have still reliably induced SAH (36,37). We verified the existence or absence of a local hemorrhage in the SAH or SAH-sham, respectively, by inspecting the MRI T2*-weighted images taken at the 90-minute post-injury time point, as described below. The criterion for exclusion was when we observed absence of hemorrhage for SAH or existence of hemorrhage in SAH-sham.

## Multi-modal Imaging and Analysis

For additional details in the description of imaging and analysis, refer to supplementary materials.

### Ultrasound and Photoacoustic Imaging

We evaluated *in vivo* longitudinal variations in cerebrovascular oxygenation (%sO$_2$), blood flow velocity, and blood perfusion in injury and sham protocols using a battery of 3D ultrasound (US), color Doppler and photoacoustic (PA) scans (Figure 1C). All US/PA protocols were performed using a Vevo 3100 LAZR-X system (FUJIFILM VisualSonics, Toronto, Canada) with a 21 MHz frequency transducer (Vevo MX 250) and the 680-970 nm laser port.

**MRI**

We acquired an MRI battery of scans to measure hemodynamic and physiological outcomes at baseline and after injury (Figure 1D). We performed all mouse brain imaging using a pre-clinical 7T MRI system (Bruker, Billerica, MA) with mouse-specific head receiver coil and ancillary equipment. We used a multi-echo RARE sequence for T2-mapping and a multi-gradient echo sequence to collect T2*-weighted images. Time of flight (TOF) angiography images were acquired using a 3D gradient echo pulse. Our diffusion tensor imaging (DTI) approach used a single shot echo planar imaging (EPI) pulse sequence. Finally, we collected three axial 2D CBF maps using a pseudo-continuous arterial spin labelling sequence (ASL) at 3 landmarks: two 1-mm slices at the bregma and one at the posterior cerebral arteries.

**Image Analysis and Post Processing**

We quantified $\%sO_2$, Trace of the diffusion tensor (measure of diffusivity) and CBF on ipsilateral and contralateral sides of injury. We drew region of interest (ROI) masks to include the 2-3 mm cortical region for maps of $\%sO_2$ and Trace and separated these masks to ipsilateral and contralateral regions at the midsagittal plane. Next, we computed the average $\%sO_2$ and Trace in each region by multiplying the mean value by the area of the 2D mask in an axial slice, summing over the slices across the brain, and dividing the resulting value by the combined area of the included regions. We used a similar approach to analyze CBF maps but summed over the three collected axial 2D images.

To analyze the color Doppler data, we divided the 3D color Doppler images into two hemispheres, using the underlaid B-mode image as the anatomical reference. In each hemisphere, we multiplied the absolute value of each voxel's measured velocity by voxel volume to obtain the velocity-weighted volume of flow (VVF) metric.

We also created 3D ROIs for pathology in T2 and Trace maps for SAH and TBI using ITK-SNAP software (38) by locating abnormal signal within the brains and quantifying its volume.

**Statistical Analysis**

For statistical analysis, we performed mixed repeated-measure analysis of variance (ANOVA) with the Greenhouse–Geisser correction to account for violations of sphericity on %$sO_2$, VVF, CBF, and Trace values, using MATLAB. The within-subject factor was time, and the between-subject factor was injury category (interaction of TBI/TBI-sham and SAH/SAH-sham with time). Given a Greenhouse-Geisser $p$ value smaller than 0.05 for the time intercept and injured/control:time interaction, we performed post-hoc Tukey-Kramer tests for multiple comparisons across time and between corresponding injury/control groups, respectively. We used partial $\eta^2$ and Hedges' g (with correction for the small sample size) as measures of effect size for the mixed repeated-measure ANOVA and multiple comparisons, respectively. Differences with $p$ values smaller than 0.05 were deemed statistically significant.

# Results

We used hypointense T2*-weighted signal as a marker for hemorrhage and excluded two animals from the SAH group due to lack of hemorrhage and one animal from the SAH-sham group due to presence of hemorrhage, likely from unintended vascular perforation.

## US/PA show qualitative oxygenation and blood flow alterations after TBI and SAH

Both TBI and SAH led to conspicuous reductions in %sO$_2$ and blood velocity that were discernable by eye on PA and color Doppler maps. The TBI-sham group did not show any abnormal pattern at any time point (Figure 2A), whereas the TBI group showed a substantial reduction in %sO$_2$ at the site of impact and across the brain 15 minutes post-injury, partially recovering at the 3-hour time point (Figure 2B). Animals that underwent arterial occlusion but not perforation (i.e., SAH-sham group) showed reduced ipsilateral %sO$_2$ and reduced PCA velocities on the ipsilateral side for both post-injury scans (Figure 2C). However, the outcomes were less severe than for the animal with vascular perforation (i.e., SAH group), which underwent a severe ipsilateral drop in %sO$_2$ with only a small change on the contralateral side at 15 minutes post injury that progressively worsened at 3 hours post injury. There was a visible reduction in velocity captured from color Doppler imaging in the ipsilateral posterior cerebral artery of the injured animals for both 15 minutes and 3 hours post injury (Figure 2D).

## Cerebral blood oxygenation decreases diffusely following TBI and SAH

Brain oxygenation (%sO$_2$) significantly decreased within whole-hemisphere ROIs on the side of injury (ipsilateral) 15 minutes after TBI (19.2% reduction, *p*<0.001, g=7.18), SAH (43.9% reduction, *p*<0.001, g=8.05), and SAH-sham (15.5% reduction, *p*<0.001, g=3.83) compared to baseline (Figure 3A, C). Oxygenation remained significantly lower than baseline for these groups 3 hours after injury as well. In the TBI group, mice showed some recovery of %sO2 from the 15-minute to the 3-hour post-injury scan (9.3% increase, *p*=0.004, g=2.26), although the 3-hour post injury ipsilateral %sO$_2$ was still 11.7% lower than baseline (*p*<0.001). Compared to the TBI-sham group, ipsilateral %sO$_2$ values for the TBI group were 20.0% (*p*<0.001, g=8.08) and 6.7% (*p*=0.012, g=1.49) lower at 15 minutes and 3 hours post injury, respectively. In the SAH group, there was no statistically significant recovery in %sO$_2$ between 15 minutes and 3 hours post injury on either side of the brain. Compared to the SAH-sham group, SAH ipsilateral %sO$_2$ values were 34.9% (*p*<0.001, g=5.19) and 25.6% (*p*=0.019, g=1.59) lower at 15 minutes and 3 hours post injury, respectively.

In addition to the %sO$_2$ values averaged over the whole-hemisphere ROIs, ipsilateral %sO$_2$ values over 2D axial slices across the brain at different time points also showed unique trends in TBI and SAH injuries. An extended part of mid-brain had reduced %sO$_2$ 15 minutes post injury in TBI (Figure 3B), while oxygenation values were close to sham levels in the anterior and posterior regions at this time point. On the other hand, the SAH %sO$_2$ pattern indicated a large deviation from sham levels in the posterior regions for both 15 minutes and 3 hours post injury scans (Figure 3D). Moreover, we did not observe any recovery in either part of the brain by the 3-hour time point.

The contralateral %sO$_2$ for the TBI group showed a reduction of 17.6% ($p<0.001$, g=2.06) 15 minutes post injury, which was also 17.68% ($p=0.002$, g=2.08) lower than TBI-sham (Figure 3E). Contralateral %sO$_2$ subsequently recovered from 15 minutes to 3 hours post-injury (11.0% increase, $p=0.0052$, g=0.93) but was still 8.5% lower than baseline ($p=0.0107$, g=1.48). However, there was no significant difference between TBI and TBI-sham at the 3-hour time point. Contralateral %sO$_2$ for the SAH group decreased by 11.6% ($p<0.001$, g=5.15) 15 minutes post-injury and was 21.2% lower than baseline ($p=0.008$, g=2.00) at 3 hours post-injury (Figure 3G). However, there was no significant difference in contralateral %sO$_2$ between SAH and SAH-sham at any time point.

Reviewing oxygenation across axial 2D slices for the contralateral side also revealed unique trends for the injury types. There was a similar post-injury dip in %sO$_2$ at 15 minutes in the mid brain region for TBI (Figure 3F). Anterior and posterior levels were close to sham levels at this time point. For SAH, however, contralateral %sO$_2$ levels were in the same range as SAH-sham 15 minutes post injury. These levels deviated from each other particularly in mid brain slices 3 hours post-injury (Figure 3H).

**Velocity-weighted volume of flow decreases following TBI and SAH**

There was a time-dependent response for all groups in the VVF measure, which was derived from color Doppler images (Figure 4). The TBI group had significant decreases compared to baseline bilaterally for both 15 minutes and 3 hours post-injury scans. Ipsilateral VVF for TBI was 41.0% ($p<0.001$, g=2.45) and 45.9% ($p<0.001$, g=1.95) lower than baseline at 15-minute and 3-hour post injury time points, respectively. The contralateral reductions for TBI with respect to baseline at the same time points were 47.5% ($p<0.001$) and 40.8% ($p<0.001$). TBI-shams had significantly different values only between baseline and 3 hours on the contralateral side (37% reduction, $p=0.002$, g=1.10). The only significant difference between TBI and TBI-sham occurred at the 15 minutes post-injury time point in the contralateral side (54.6%, $p=0.039$, g=1.15).

For the SAH and SAH-sham groups, on the ipsilateral side, VVF significantly decreased at all post injury time points compared to baseline. Ipsilateral VVF for SAH was 85.0% lower than baseline at 15-minute ($p<0.001$, g=7.31) and 93.3% lower than baseline at 3-hour ($p<0.001$, g=12.6) post injury. Ipsilateral VVF for SAH-sham decreased by 55.7% ($p<0.001$, g=2.72) from baseline to 15 minutes post injury and again by 55.2% ($p=0.004$, g=1.87) from 15 minutes to 3 hours post injury. Contralateral VVF for SAH was 52.2% ($p=0.045$, g = 4.03) and 71.1% ($p=0.025$, g=2.48) lower than baseline at 15-minute and 3-hour post injury time points, respectively. Three hours after SAH-sham, contralateral VVF reduced by 47.7% ($p=0.039$, g=1.23) and 44.3% ($p=0.002$, g=0.81) compared to baseline and 15-minute post injury time points. There were no significant differences in VVF when comparing SAH and SAH-sham.

## MRI markers show qualitative focal and diffuse hemorrhage, edema, and ischemia after TBI and SAH

A summary of visual observations in MR results can be found in Figure 5A. The baseline scans did not show any abnormalities.

Following TBI, we generally observed local T2*-weighted (hemorrhage), T2, Trace, and CBF abnormalities at the post-injury MRI scans. The TBI-sham group did not have any signs of pathology in any of the scans (Figure 5B). On the other hand, in a majority (5/6) of the TBI mice, we observed a focal region with low T2 and Trace values and T2*-weighted hypointensities in a core region near the impact site surrounded by small adjacent regions of increased T2 and Trace (Figure 5C). ASL maps showed reduced CBF near the CCI site (4/6) or diffusely (2/6) following TBI while TOF maps did not indicate any visible change in vessel intensity in the major arterial vessels. Additional abnormalities appearing less frequently included focal increase in T2, T2*-weighted, and Trace (1/6, Figure 6A), abnormalities distant from the impact site in the ventral part of the brain near the olfactory bulb (2/6, Figure 6B) and contusions surrounded by vasogenic peripheries at several different cortical locations (1/6, Figure 6C).

The SAH-sham group did not have any T2, T2*-weighted or Trace abnormalities (Figure 5D) but did exhibit ipsilateral CBF reductions and reduced signal intensity in the CCA on TOF images suggesting successful occlusion in the SAH-sham group. All of the remaining animals in the SAH group (4/4) exhibited hypointense T2*-weighted signal within the brain (Figure 5E and 6D-F) and varying extents of increased T2 and reduced Trace values. ASL maps showed reduced CBF for all animals (4/4), extending bilaterally in three of four animals. TOF scans for all four SAH cases visibly showed reduced vessel intensity in the ipsilateral CCA.

**Abnormal T2 and Trace regions are more focal for TBI than SAH**

Visually, SAH appeared to affect a more extensive region of the brain, while MRI abnormalities after TBI were focal. To assess this observation quantitatively, we calculated the volumetric extent of abnormal T2 and Trace values for the two injury types (Figure 7A). The extent of regions with increased T2 and decreased Trace following SAH were 16.58±24.43 mm$^3$ and 13.44±12.35 mm$^3$, respectively. Following TBI, increased T2 was observed with extent of 2.23±1.23 mm$^3$ while increased Trace had an extent of 1.74±1.57 mm$^3$.

**CBF values were bilaterally reduced following SAH but not TBI**

TBI and TBI-sham did not show any significant changes between or within groups for hemispheric ROI values of CBF measured by ASL. On the other hand, there were bilateral decreases in CBF averaged over the three ASL ipsilateral and contralateral slices post-injury in the SAH group compared to baseline (Figure 7B). Post-injury Ipsilateral and contralateral CBF values for SAH were 71.4% ($p<0.001$, g=2.98) and 52.2% ($p=0.009$, g=1.65) lower than baseline. The SAH-sham group also had a 45.9% decrease ($p=0.005$, g=3.38) on the ipsilateral side compared to baseline. The ipsilateral CBF in the SAH group was 34.8% lower than SAH-sham post injury ($p=0.036$, g=1.36).

**Hemispheric diffusivity decreased following SAH but not TBI**

There was no significant difference in Trace for hemispheric ROI values between TBI and TBI-sham, nor was there any difference for either group between baseline and post-injury values. Conversely, post-injury ipsilateral and contralateral Trace were 17.2% ($p$=0.001, g=1.42) and 6.9% ($p$=0.003, g=1.51) lower than baseline for the SAH group (Figure 7C). Comparing SAH and SAH-sham post-injury values, the SAH group had significantly lower Trace compared to SAH-sham on the contralateral side (2.9% reduction, $p$=0.015, g=1.68). However, we did not observe any significant difference between SAH and SAH-sham for the ipsilateral side post injury.

## Discussion

We characterized the hyperacute vascular and tissue responses to two related experimental injuries with cerebrovascular involvement (TBI and SAH), as early as 15 minutes up to 3 hours post-injury using a combination of color Doppler, photoacoustics, and MRI. The primary photoacoustic findings for TBI were extensive and bilateral decreased %$sO_2$ at 15 minutes post injury with partial recovery 3 hours post injury. The main MRI outcomes after TBI indicated focal T2, T2*-weighted, and Trace abnormalities, possibly due to hemorrhage and edema pathology and more extensive perfusion deficit post injury. In the case of SAH, cerebral %$sO_2$ was reduced bilaterally following SAH, but more severely on the ipsilateral side 15 minutes post injury and, unlike TBI, %$sO_2$ remained significantly lower than baseline even 3 hours post injury. Abnormal biomarkers from MRI outcomes were more prominent and extensive including focal hemorrhage with varying extents of edema and diffuse ipsilateral perfusion deficit.

## Hemorrhage and Edema

Focal hemorrhage and the consequent degradation of blood products are hallmarks of physical injury to the tissue. Deoxyhemoglobin is strongly paramagnetic due to its four unpaired electrons and thus has a strong T2*-weighted relaxation shortening effect (39). Therefore, hemorrhage and onset of hematoma are associated with T2*-weighted signal loss and hypointense regions in the T2*-weighted scans. Human MRI observations in the acute phase support the existence of a hemorrhagic core in contusion injuries (20) while acute clinical (40,41) and hyperacute preclinical studies (36,42) report hemorrhage and hematoma outcomes in SAH. The T2*-weighted images in our study revealed focal hemorrhage in both TBI and SAH models. The majority of the TBIs had a focal hemorrhagic core at the site of impact in the cortex while the focal hemorrhage for the SAH model was deeper inside the brain and was likely due to blood flooding into perivascular spaces and pooling into the parenchyma.

The edema pathology coexisting with this hemorrhagic injury was also uniquely different between the TBI and SAH models. Formation of edema is a common early outcome of TBI and SAH. Cytotoxic edema, which is associated with cellular swelling that can be induced by chemical or mechanical factors, occurs as cell volume changes due to intracellular transfer of water and disruption of ionic gradients through the cell membrane. Vasogenic edema, on the other hand, is associated with increased vascular permeability, which can lead to the extravasation of blood serum proteins in the extracellular space following blood brain barrier disruption. In principle, cytotoxic edema eventually leads to cell death and is irreversible, while vasogenic edema might be due to transient inflammatory processes and is reversible

(43). Due to their close relationship, usually both types of edema are present in some degree following a brain insult (44,45).

MRI observations rely on diffusivity changes as biomarkers for edema. In this context, enhanced and restricted diffusion can be interpreted as vasogenic and cytotoxic edema, respectively. A clinical study revealed the formation of a vasogenic edema rim surrounding the contusion core, although the earliest scans belonged to 50 hours post-injury (20). Mouse and rat studies of open skull CCI have reported vasogenic edema formation as early as 1h post injury supported by T2 and ADC maps but without a hypointense T2/ADC core (27,46,47). Our current results from a closed-head moderate TBI also show the formation of vasogenic edema around a hemorrhagic contusion core for the TBI model.

On the other hand, the edema resulting from SAH is initially cytotoxic in nature. Clinical studies have shown the existence of cytotoxic edema in the acute phase following SAH (39,40), reporting that vasogenic edema is a later outcome of it (39). Our SAH results demonstrate the cytotoxic edema pathology in the hyperacute stage after vascular rupture. As opposed to TBI, edema outcome after SAH was not limited to the area immediately surrounding the hemorrhage and was spatially variable.

## Perfusion, Hemodynamics, and Oxygenation

Reduced cerebral perfusion could be due to either local tissue damage and subsequent microvascular dysfunction or a decrease in global blood supply to the brain. We observed focal reductions in tissue perfusion in our TBI model. These low-perfusion regions were larger than the hemorrhagic core of impact and the surrounding edema rim. Although there is coupling between local CBF and neuronal metabolism through the neurovascular unit, primary (physical tissue damage) and secondary (e.g., blood-brain barrier disruption) injuries can interfere with normal neurovascular operation and consequently decrease local CBF (1,48). A recently proposed idea suggests an early outcome of TBI is the formation of a traumatic low-perfusion penumbra around the contusion core in focal lesions (49). The kinetic energy absorbed during the impact excessively deforms the microvessels, resulting in the contusion core, while the surrounding penumbra absorbs energy to a smaller extent (50) but still enough to compromise blood brain barrier permeability and lead to vasogenic edema (51). Previous human and animal studies have shown that this expanded area has reduced perfusion in the acute phase and its size correlates with the extent of necrotic tissue determined at later follow-up observations (52,53). This hyperacute pathological mechanism and its relationship with possible neuronal changes and neurovascular dysregulation require further research.

Injury also alters blood flow velocities in the brain. One of the most used non-invasive methods to measure local velocity in the brain is transcranial ultrasound. In a clinical study, low velocities in the MCA, acquired upon admission of mild to moderate TBI patients, were associated with later neurologic deterioration (54). On the other hand, local MCA blood velocity increases measured using transcranial ultrasound occur with vasospasm following SAH (55,56). However, a global measure of blood flow velocity in the brain has not been used in the literature. We used color Doppler to measure blood flow velocity in the major vessels and extracted velocity-weighted volume of flow (VVF). The VVF results showed longitudinal reductions for both TBI and SAH, with minimal velocity signal on the ipsilateral side for SAH. The ipsilateral VVF for both SAH and SAH-sham showed reductions compared to baseline, which was expected given the permanent ligation of ECA and CCA for both groups, as well as the hemorrhagic injury in the SAH group.

Our results revealed the diffuse reduction in tissue level perfusion as a hyperacute outcome of SAH. Considering perfusion and VVF results together allows us to understand the relationship with large vessel hemodynamics and tissue level perfusion. Pre-post comparisons for both CBF and VVF show significant reductions after the injury for SAH and SAH-sham. However, while both SAH and SAH-sham have VVF reductions, their post-injury VVFs are not statistically different from each other, and only the ipsilateral perfusion in SAH is lower than SAH-sham. In other words, the surgery and sham operations both affect major vessel hemodynamics. However, the hemorrhagic event results in significantly lower tissue-level perfusion in SAH compared to SAH-sham. A plausible explanation for this trend is the eventual global increase in intracranial pressure due to SAH. Once the subarachnoid space and subsequently the ventricles are filled up by blood on the ipsilateral side, there will be extravasation to the contralateral hemisphere after a few hours. It is at this time point that we see a global perfusion deficit for SAH, which involves hemorrhage, and not SAH-sham, which only involves CCA ligation. Moreover, the hyperacute response after cerebral hemorrhage has been more associated with microcirculatory dysfunction in the form of microthrombosis and blood-brain barrier damage, whereas large artery spasms are commonly observed days after the initial vascular rupture (57).

Decrease in $\%sO_2$ after TBI is a common secondary injury. It may indicate a delay in energy-dependent recovery due to unbalanced oxygen consumption and delivery temporarily disrupted by the mechanical injury (58). Clinical studies show that TBI patients with systemic hypoxia prior to or on admission to hospital have higher mortality and poorer outcomes than patients with normal %sO$_2$ levels (59,60). Our study is the first to quantify the reduction in arterial blood %sO$_2$ in the brain as early as 15 minutes after the insult.

Review of the literature reveals possible explanations for the immediate global decrease and subsequent recovery of cerebral %sO$_2$ in our moderate TBI model. The immediate response to TBI is associated with reduced tissue-level perfusion (26,27) and increased cerebral metabolism of glucose (28,61,62). Both events could explain the global decrease in %sO$_2$ 15 minutes post-TBI in our study. Cerebral metabolism of glucose recovers 1-6 hours post injury and goes into a depression period afterwards that could take up to 10 days (28). The early turning point in metabolism trend may explain the normal global values seen in our 3 hours post injury final %sO$_2$ scans. It is worth noting that another study (23) tracked focal %sO$_2$ values after TBI at baseline and 6 hours post injury and showed statistically significant difference between these two times. They saw recovery to baseline values at day 1 post injury. In our study, The TBI group still had significantly reduced ipsilateral %sO$_2$ compared to both its own baseline and TBI-sham at 3 hours post injury. However, their injury model was juvenile closed-head mild TBI and they computed %sO$_2$ only at the left somatosensory cortex, where the contusion happened, whereas we used adult mice and averaged %sO$_2$ over the brain hemisphere.

The SAH %sO$_2$ decrease in our study was highly lateralized and persisted throughout the scanning timeline of our experiment. The SAH post-injury ipsilateral %sO$_2$ values were smaller than SAH-sham, demonstrating the effect of vascular rupture, hemorrhage and the likely reduction in perfusion pressure against mere ligation of supplying vessels in the sham procedure. Lack of significant contralateral difference between the SAH and SAH-sham %sO$_2$ responses over time indicates the highly lateralized outcome of this type of hemorrhagic injury.

## Limitations and Future Work

Our study had several limitations. (1) The animals had to remain under anesthesia for at least four hours on the injury day. The prolonged anesthesia potentially affects some of the physiological metrics that we tracked, as evident in the %sO$_2$ and VVF trends for the TBI-sham group. However, this period was the shortest possible duration to do the injury and the relatively comprehensive hyperacute imaging until three hours post-injury. (2) Signal to noise ratio decreases with increasing depth in photoacoustic imaging. That is why we mainly drew our %sO$_2$ ROIs down to 2 mm below the skull and averaged the values in this ROI. Improvements in PA imaging could allow resolving these deeper parts of the brain in the future.

Future directions following this research could include: (1) acquiring US/PA scans with shorter intervals in the hyperacute phase post injury, which allows tracking the time course of cerebral %sO$_2$ and VVF response and identifying key times in evolution/recovery of injury, and (2) developing methods to co-register US/PA and MR images, which enables us to perform voxel-wise analysis of outcomes using biomarkers from both modalities.

## Acknowledgements

The work was supported by the National Institutes of Health (NIH) National Heart, Lung, and Blood Institute (NHLBI) grant number R00HL140106, and National Institute of Biomedical Imaging and Bioengineering (NIBIB) Trailblazer award number R21EB032187, as well as AARGD-21-850835 from the Alzheimer's Association. All imaging was performed in the UA Translational Bioimaging Resource (TBIR) and made possible by support from the Research, Innovation & Impact (RII), the Technology Research Initiative Fund, and the NIH small instrumentation grant – S10 OD025016. We thank Christy Howison for technical support during MRI scanning.

## Author Contribution Statement

AK and LD contributed to study design, performed the imaging and the TBI model, analyzed the results, and co-wrote the draft. ECP performed the SAH surgery and contributed to MR ROI analysis and draft preparation. CAP contributed to co-localizing MR and US/PA images, analyzing CBF maps, and draft preparation. RSW contributed to study design, optimized the parameters for US/PA scans, and revised the draft. PWP, EBH, and KL designed the study, supervised the experiments and analysis, and contributed to writing and revising the draft.

## Conflict of Interest Statement

The authors declares that there is no conflict of interest.

# References

1. Kenney K, Amyot F, Haber M, Pronger A, Bogoslovsky T, Moore C, et al. Cerebral vascular injury in traumatic brain injury. Exp Neurol. 2016;275:353–66.

2. Chandra A, Stone CR, Du X, Li WA, Huber M, Bremer R, et al. The cerebral circulation and cerebrovascular disease III: Stroke. Brain Circ. 2017;3(2):66.

3. Deshpande A, Jamilpour N, Jiang B, Michel P, Eskandari A, Kidwell C, et al. Automatic segmentation, feature extraction and comparison of healthy and stroke cerebral vasculature. NeuroImage Clin. 2021;30(March 2020):102573.

4. Macdonald RL, Schweizer TA. Spontaneous subarachnoid haemorrhage. Lancet. 2017;389(10069):655–66.

5. Meaney DF, Smith DH. Cellular biomechanics of central nervous system injury. 1st ed. Vol. 127, Handbook of clinical neurology. Elsevier B.V.; 2015. 105–114 p.

6. Sato M, Chang E, Igarashi T, Noble LJ. Neuronal injury and loss after traumatic brain injury: time course and regional variability. Brain Res. 2001;917(1):4554.

7. Ziebell JM, Taylor SE, Cao T, Harrison JL, Lifshitz J. Rod microglia: elongation, alignment, and coupling to form trains across the somatosensory cortex after experimental diffuse brain injury. J Neuroinflammation. 2012;

8. Tong W, Zheng P, Xu J, Guo Y, Zeng J, Yang W, et al. Early CT signs of progressive hemorrhagic injury following acute traumatic brain injury. Neuroradiology. 2011 May;53(5).

9. Lawrence TP, Pretorius PM, Ezra M, Cadoux-Hudson T, Voets NL. Early detection of cerebral microbleeds following traumatic brain injury using MRI in the hyper-acute phase. Neurosci Lett. 2017;655:143–50.

10. Jha RM, Kochanek PM, Simard JM. Pathophysiology and treatment of cerebral edema in traumatic brain injury. Neuropharmacology. 2019 Feb;145.

11. Shiozaki T, Hayakata T, Tasaki O, Hosotubo H, Fuijita K, Mouri T, et al. CEREBROSPINAL FLUID CONCENTRATIONS OF ANTI-INFLAMMATORY MEDIATORS IN EARLY-PHASE SEVERE TRAUMATIC BRAIN INJURY. Shock. 2005 May;23(5).

12. Hall ED, Bryant Y, Cho W, Sullivan PG. Evolution of post-traumatic neurodegeneration after controlled cortical impact traumatic brain injury in mice and rats as assessed by the de Olmos silver and fluorojade staining methods. J Neurotrauma. 2008;25(3):235–47.

13. Stoica BA, Faden AI. Cell death mechanisms and modulation in traumatic brain injury. Neurotherapeutics. 2010;7(1):3–12.

14. Hill-Felberg SJ, McIntosh TK, Oliver DL, Raghupathi R, Barbarese E. Concurrent loss and proliferation of astrocytes following lateral fluid percussion brain injury in the adult rat. J Neurosci Res. 1999;57(2):271279.

15. van Lieshout JH, Dibué-Adjei M, Cornelius JF, Slotty PJ, Schneider T, Restin T, et al. An introduction to the pathophysiology of aneurysmal subarachnoid hemorrhage. Neurosurg Rev. 2018 Oct;41(4).

16. Dóczi T, Nemessányi Z, Szegváry Z, Huszka E. Disturbances of Cerebrospinal Fluid Circulation during the Acute Stage of Subarachnoid Hemorrhage. Neurosurgery. 1983 Apr;12(4).

17. Armin SS, Colohan ART, Zhang JH. Traumatic subarachnoid hemorrhage: our current understanding and its evolution over the past half century. Neurol Res. 2006;28(4):445–52.

18. Meislin H, Criss EA, Judkins D, Berger R, Conroy C, Parks B, et al. Fatal trauma: The modal distribution of time to death is a function of patient demographics and regional resources. J Trauma - Inj Infect Crit Care. 1997 Sep;43(3):433–40.

19. Alexis R, Jagdish S, Sukumar S, Pandit V, Palnivel C, Antony M. Clinical profile and autopsy findings in fatal head injuries. J Emergencies, Trauma Shock. 2018 Jul;11(3):205–10.

20. Newcombe VFJ, Williams GB, Outtrim JG, Chatfield D, Abate MG, Geeraerts T, et al. Microstructural basis of contusion expansion in traumatic brain injury: insights from diffusion tensor imaging. J Cereb Blood Flow Metab. 2013;33(6):855–62.

21. Ljungqvist J, Nilsson D, Ljungberg M, Sörbo A, Esbjörnsson E, Eriksson-Ritzén C, et al. Longitudinal study of the diffusion tensor imaging properties of the corpus callosum in acute and chronic diffuse axonal injury. Brain Inj. 2011 Apr 1;25(4):370–8.

22. Jullienne A, Obenaus A, Ichkova A, Savona-Baron C, Pearce WJ, Badaut J. Chronic cerebrovascular dysfunction after traumatic brain injury. J Neurosci Res. 2016 Jul

1;94(7):609–22.

23. Ichkova A, Rodriguez-Grande B, Zub E, Saudi A, Fournier ML, Aussudre J, et al. Early cerebrovascular and long-term neurological modifications ensue following juvenile mild traumatic brain injury in male mice. Neurobiol Dis. 2020;141(February):104952.

24. Bouzat P, Oddo M, Payen JF. Transcranial doppler after traumatic brain injury: Is there a role? Curr Opin Crit Care. 2014;20(2):153–60.

25. Lawrence T. Imaging biomarkers in the hyper-acute phase following traumatic brain injury. University of Oxford; 2018.

26. Villapol S, Balarezo MG, Affram K, Saavedra JM, Symes AJ. Neurorestoration after traumatic brain injury through angiotensin II receptor blockage. Brain. 2015 Nov 1;138(11):3299–315.

27. Long JA, Watts LT, Li W, Shen Q, Muir ER, Huang S, et al. The effects of perturbed cerebral blood flow and cerebrovascular reactivity on structural MRI and behavioral readouts in mild traumatic brain injury. J Cereb Blood Flow Metab. 2015 Jun 24;35(11):1852–61.

28. Yoshino A, Hovda DA, Kawamata T, Katayama Y, Becker DP. Dynamic changes in local cerebral glucose utilization following cerebral concussion in rats: evidence of a hyper-and subsequent hypometabolic state. Brain Res. 1991;561(1):106–19.

29. Pasco A, Lemaire L, Franconi F, Lefur Y, Noury F, Saint-André J-P, et al. Perfusional deficit and the dynamics of cerebral edemas in experimental traumatic brain injury using perfusion and diffusion-weighted magnetic resonance imaging. J Neurotrauma. 2007;24(8):1321–30.

30. Ren Z, Iliff JJ, Yang L, Yang J, Chen X, Chen MJ, et al. 'Hit & Run'' model of closed-skull traumatic brain injury (TBI) reveals complex patterns of post-traumatic AQP4 dysregulation.' J Cereb Blood Flow Metab. 2013 Jun;33(6):834.

31. Wilding LA, Hampel JA, Khoury BM, Kang S, Machado-Aranda D, Raghavendran K, et al. Benefits of 21% oxygen compared with 100% oxygen for delivery of isoflurane to mice (Mus musculus) and rats (Rattus norvegicus). J Am Assoc Lab Anim Sci. 2017;56(2):148–54.

32. Blevins CE, Celeste NA, Marx JO. Effects of Oxygen Supplementation on Injectable and Inhalant Anesthesia in C57BL/6 Mice. J Am Assoc Lab Anim Sci. 2021;60(3):289–97.

33. Griffiths BB, Sahbaie P, Rao A, Arvola O, Xu L, Liang D, et al. Pre-treatment with microRNA-181a antagomir prevents loss of parvalbumin expression and preserves novel object recognition following mild traumatic brain injury. Neuromolecular Med. 2019;21(2):170–81.

34. Rodriguez-Grande B, Obenaus A, Ichkova A, Aussudre J, Bessy T, Barse E, et al. Gliovascular changes precede white matter damage and long-term disorders in juvenile mild closed head injury. Glia. 2018;66(8):1663–77.

35. Schüller K, Bühler D, Plesnila N. A murine model of subarachnoid hemorrhage. J Vis Exp. 2013 Nov 21;(81):e50845–e50845.

36. Wang Z, Chen J, Toyota Y, Keep RF, Xi G, Hua Y. Ultra-Early Cerebral Thrombosis Formation After Experimental Subarachnoid Hemorrhage Detected on T2* Magnetic Resonance Imaging. Stroke. 2021;1033–42.

37. Kamii H, Kato I, Kinouchi H, Chan PH, Epstein CJ, Akabane A, et al. Amelioration of Vasospasm After Subarachnoid Hemorrhage in Transgenic Mice Overexpressing CuZn–Superoxide Dismutase. Stroke. 1999 Apr;30(4):867–72.

38. Yushkevich PA, Piven J, Hazlett HC, Smith RG, Ho S, Gee JC, et al. User-guided 3D active contour segmentation of anatomical structures: significantly improved efficiency and reliability. Neuroimage. 2006;31(3):1116–28.

39. Weimer JM, Jones SE, Frontera JA. Acute cytotoxic and vasogenic edema after subarachnoid hemorrhage: a quantitative MRI study. Am J Neuroradiol. 2017;38(5):928–34.

40. Schubert GA, Seiz M, Hegewald AA, Manville J, Thomé C. Hypoperfusion in the acute phase of subarachnoid hemorrhage. Early Brain Inj or Cereb Vasospasm. 2011;35–8.

41. Bradley Jr WG. MR appearance of hemorrhage in the brain. Radiology. 1993;189(1):15–26.

42. Rumboldt Z, Kalousek M, Castillo M. Hyperacute subarachnoid hemorrhage on T2-weighted MR images. Am J Neuroradiol. 2003;24(3):472–5.

43. Hudak AM, Peng L, Marquez de la Plata C, Thottakara J, Moore C, Harper C, et al. Cytotoxic and vasogenic cerebral oedema in traumatic brain injury: assessment with FLAIR and DWI imaging. Brain Inj. 2014;28(12):1602–9.

44. Klatzo I. Pathophysiological aspects of brain edema. Acta Neuropathol. 1987;72(3):236–9.

45. Michinaga S, Koyama Y. Pathogenesis of brain edema and investigation into anti-edema drugs. Int J Mol Sci. 2015;16(5):9949–75.

46. Long JA, Watts LT, Chemello J, Huang S, Shen Q, Duong TQ. Multiparametric and longitudinal MRI characterization of mild traumatic brain injury in rats. J Neurotrauma. 2015;32(8):598–607.

47. Lu H, Lei X. The apparent diffusion coefficient does not reflect cytotoxic edema on the uninjured side after traumatic brain injury. Neural Regen Res. 2014;9(9):973.

48. Pop V, Badaut J. A neurovascular perspective for long-term changes after brain trauma. Transl Stroke Res. 2011;2(4):533–45.

49. Adatia K, Newcombe VFJ, Menon DK. Contusion progression following traumatic brain injury: a review of clinical and radiological predictors, and influence on outcome. Neurocrit Care. 2021;34:312–24.

50. Kurland D, Hong C, Aarabi B, Gerzanich V, Simard JM. Hemorrhagic progression of a contusion after traumatic brain injury: A review. J Neurotrauma. 2012;29(1):19–31.

51. Martínez-Valverde T, Vidal-Jorge M, Martínez-Saez E, Castro L, Arikan F, Cordero E, et al. Sulfonylurea receptor 1 in humans with post-traumatic brain contusions. J Neurotrauma. 2015;32(19):1478–87.

52. Von Oettingen G, Bergholt B, Gyldensted C, Astrup J. Blood flow and ischemia within traumatic cerebral contusions. Neurosurgery. 2002;50(4):781–90.

53. Plesnila N, Friedrich D, Eriskat J, Baethmann A, Stoffel M. Relative cerebral blood flow during the secondary expansion of a cortical lesion in rats. Neurosci Lett. 2003;345(2):85–8.

54. Bouzat P, Almeras L, Manhes P, Sanders L, Levrat A, David J-S, et al. Transcranial Doppler to predict neurologic outcome after mild to moderate traumatic brain injury. Anesthesiology. 2016;125(2):346–54.

55. Fontanella M, Valfrè W, Benech F, Carlino C, Garbossa D, Ferrio M, et al. Vasospasm after SAH due to aneurysm rupture of the anterior circle of Willis: value of TCD monitoring. Neurol Res. 2008;30(3):256–61.

56. Frontera JA, Fernandez A, Schmidt JM, Claassen J, Wartenberg KE, Badjatia N, et al. Defining vasospasm after subarachnoid hemorrhage: what is the most clinically relevant definition? Stroke. 2009;40(6):1963–8.

57. Terpolilli NA, Brem C, Bühler D, Plesnila N. Are we barking up the wrong vessels? Cerebral microcirculation after subarachnoid hemorrhage. Stroke. 2015;46(10):3014–9.

58. Martini RP, Deem S, Treggiari MM. Targeting brain tissue oxygenation in traumatic brain injury. Respir Care. 2013;58(1):162–72.

59. Spaite DW, Hu C, Bobrow BJ, Chikani V, Barnhart B, Gaither JB, et al. The Effect of Combined Out-of-Hospital Hypotension and Hypoxia on Mortality in Major

Traumatic Brain Injury. Ann Emerg Med. 2017;69(1):62–72.

60. McHugh GS, Engel DC, Butcher I, Steyerberg EW, Lu J, Mushkudiani N, et al. Prognostic value of secondary insults in traumatic brain injury: Results from the IMPACT study. J Neurotrauma. 2007;24(2):287–93.

61. Giza CC, Hovda DA. The new neurometabolic cascade of concussion. Neurosurgery. 2014;75(suppl_4):S24–33.

62. Giza CC, Hovda DA. The neurometabolic cascade of concussion. J Athl Train. 2001;36(3):228.

# Figures

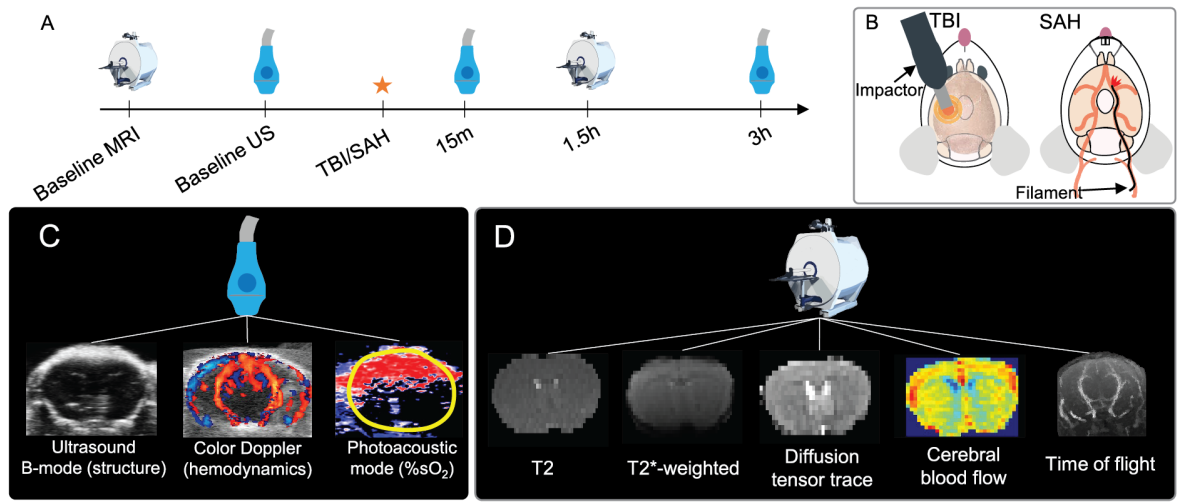

**Figure 1.** Overview of the experimental design and utilized imaging modalities: magnetic resonance imaging (MRI) and ultrasound (US). A) Timeline of experiments and imaging sessions. B) Schematics of closed head traumatic brain injury (TBI) and subarachnoid hemorrhage (SAH). C,D) Breakdown of multi-modal imaging approach.

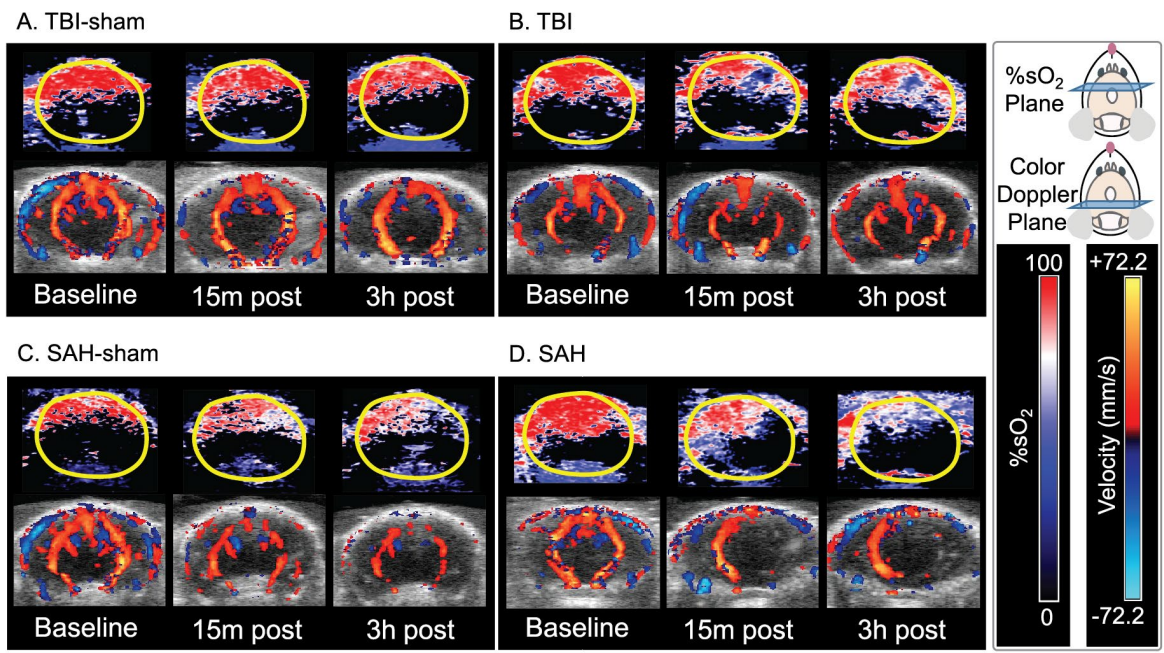

**Figure 2.** Cerebral blood oxygen saturation (%sO2) and color Doppler results for the A) TBI-sham, B) TBI, C) SAH-sham, and D) SAH cohorts. A) TBI-sham does not have any visible changes in either scan. B) The TBI images show a severe focal decrease in oxygenation at the impact site 15 minutes post injury that partially recovers by the 3-hour time point. C) The SAH-sham images show a slight ipsilateral reduction in oxygenation and blood velocity 15 minutes post injury that worsens at the 3-hour post-injury scan. D) SAH results show a lateralized severe decrease in oxygenation accompanied by absence of blood velocities captured by the device in the ipsilateral side 15 minutes following the injury, and the oxygenation response deteriorates further 3 hours post injury. Color Doppler images were produced by computing the maximum intensity projection of negative and positive velocity values over 8 axial slices (amounting to 1.2 mm) centered at the branching point of posterior cerebral arteries.

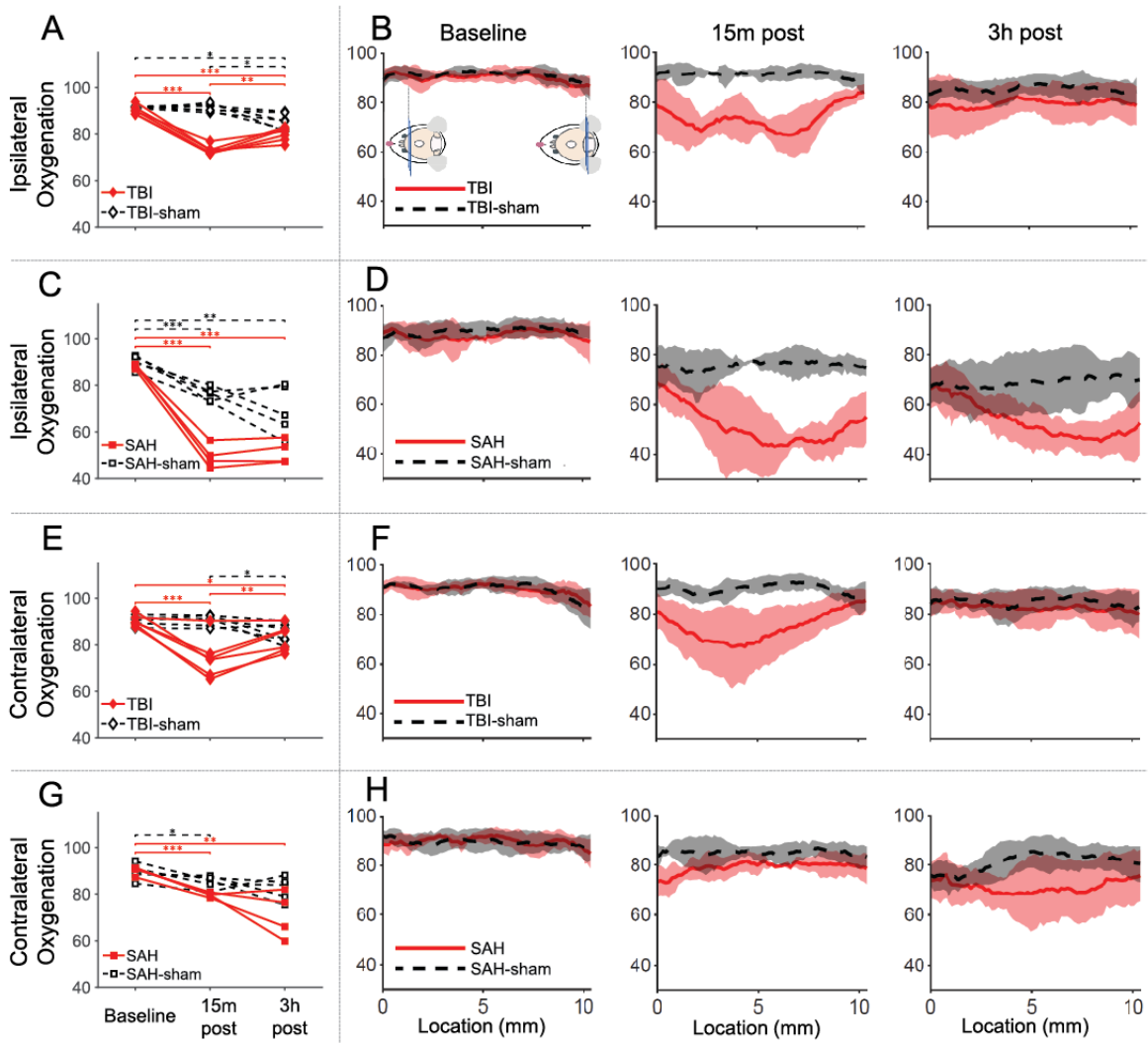

**Figure 3.** A, C, E, G) Mean ipsilateral brain blood oxygen saturation (%sO2) averaged over the brain on ipsilateral and contralateral sides (* p < .05, ** p < .01, *** p < .001). B,D,F,H) Ipsilateral and contralateral %sO2 per axial slices across 10.5 mm of brain starting from 3 mm anterior to bregma to the posterior part of the brain (solid line: mean oxygenation at slice, shade: standard deviation). A) Ipsilateral %sO2 decreases significantly in TBI followed by a partial recovery. B) Ipsilateral %sO2 values are mainly lower than TBI-sham in anterior to mid-brain, which is also in the vicinity of impact. C) Ipsilateral %sO2 decreases significantly in SAH without recovery. D) The difference between ipsilateral %sO2 in SAH and SAH-sham is in the mid to posterior region. E) Mean contralateral brain blood oxygenation trends

show a severe decrease for all but one TBI animal on the contralateral side followed by a partial recovery. F) TBI contralateral oxygenation values are lower than TBI-sham mostly in mid-brain and are close to sham values in anterior and posterior regions. G) The contralateral reduction for the SAH group is less severe than ipsilateral and in the same order as SAH-sham. H) Trends of SAH and SAH-sham across the brain are similar to each other 15 minutes post injury. These two trends slightly diverge 3 hours following the injury when SAH exhibits reduced oxygenation mostly in mid-brain.

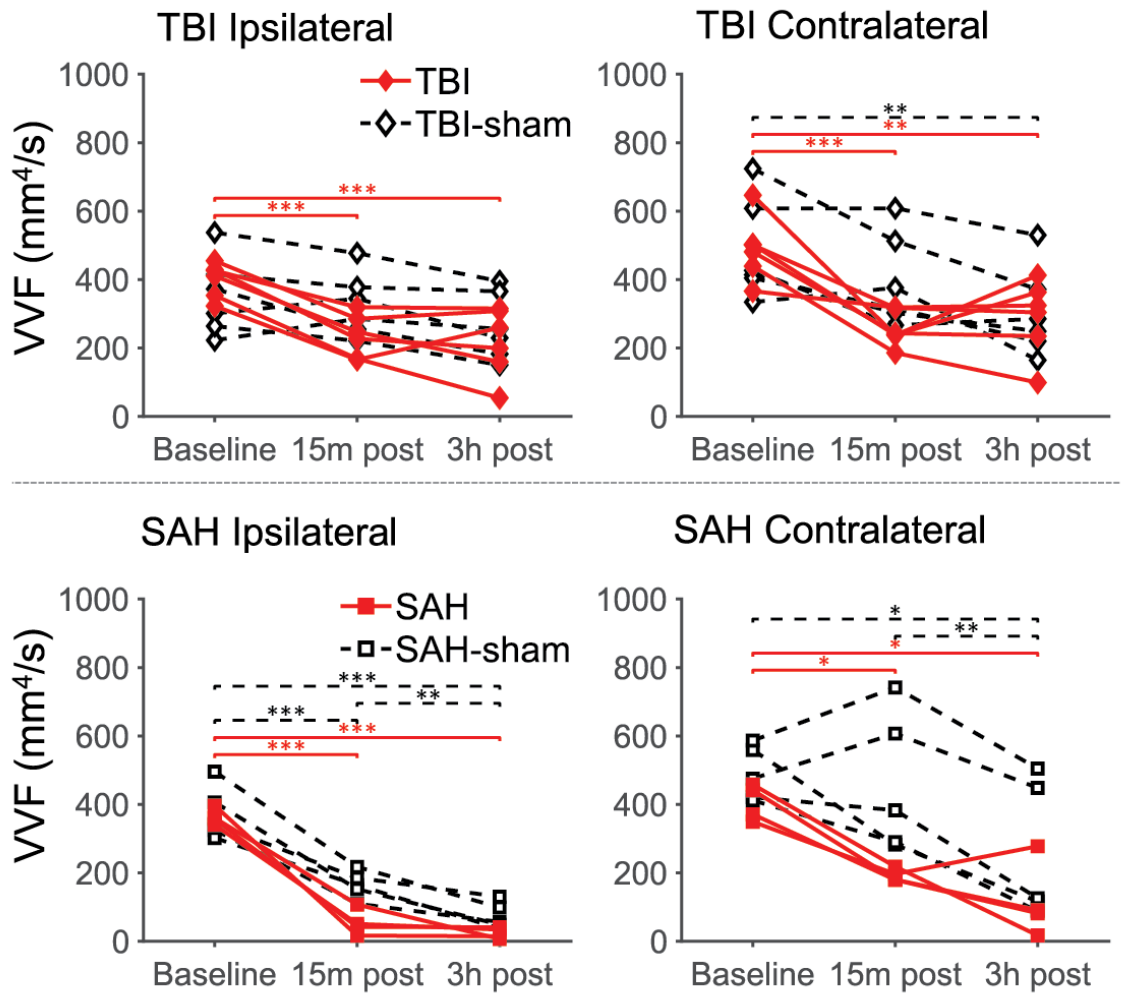

**Figure 4.** Trends of velocity-weighted volume of flow (VVF) from color Doppler images (* p < .05, ** p < .01, *** p < .001). The results show a decreasing trend from baseline to 3 hours post injury for all study groups, but the SAH and SAH-sham groups both have the most severe ipsilateral drops in VVF.

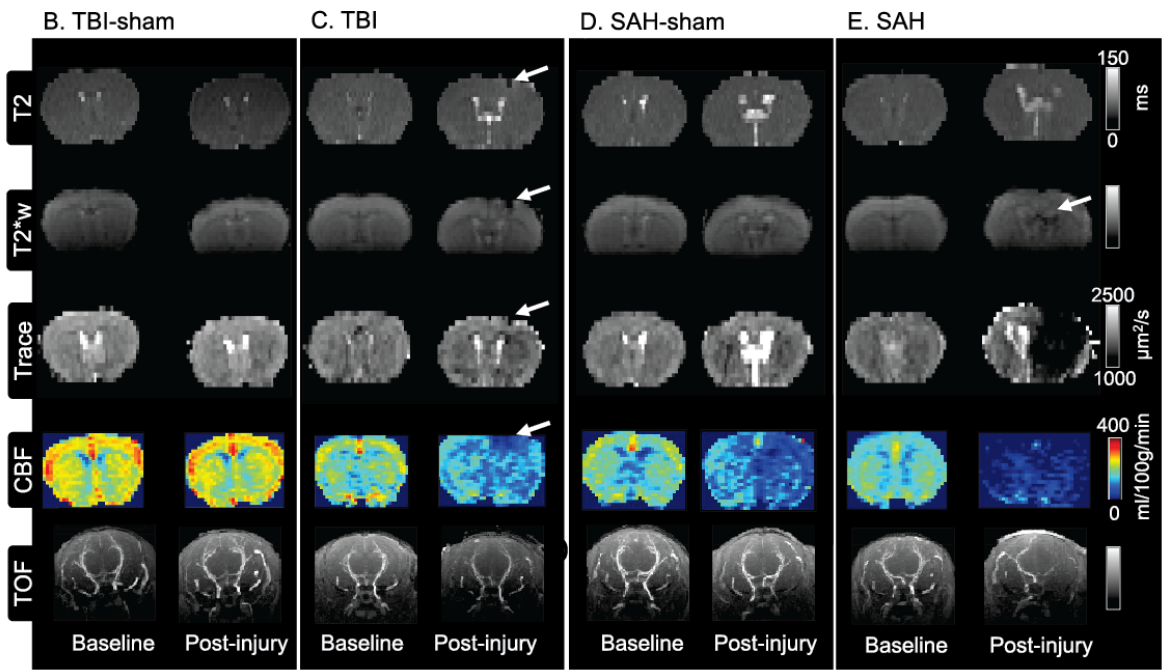

**Figure 5.** A) Magnetic resonance imaging (MRI) summary of findings and B) images detailing the T2 relaxation, T2*-weighted, diffusion tensor Trace, cerebral blood flow (CBF) obtained via arterial spin labeling (ASL), and time-of-flight (TOF) angiography, all acquired at baseline and 90 m following the injury or sham procedure for the B) TBI-sham, C) TBI, D) SAH-sham, and E) SAH cohorts. B and D) In the sham groups, the only visible pathology is the ipsilateral drop in CBF for SAH-sham. C) The TBI injury shows a focal hypointense region in T2, T2*W, and Trace maps under the impact site indicating hemorrhage. There is a small hyperintense T2 and Trace region enclosing the hemorrhagic core indicating vasogenic edema. The CBF map also shows a focal reduction at this impact site. E) In addition to the small elongated hemorrhagic site visible in the T2*-weighted image, an expanded hyperintense T2 and hypointense Trace region in the post-injury scans for the SAH group indicate cytotoxic edema. CBF is decreased bilaterally and the TOF image shows reduced vessel intensity on the ipsilateral side.

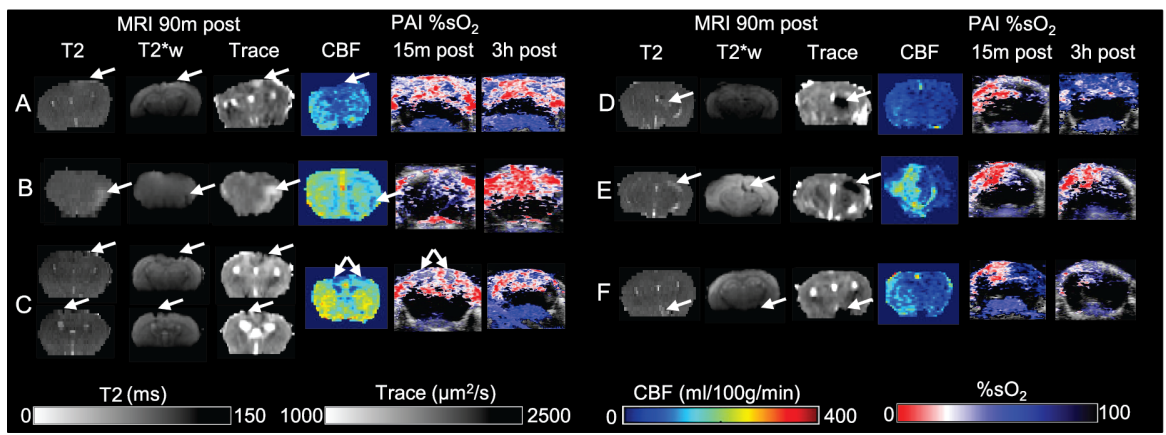

**Figure 6.** Spatially distinct MRI and oxygenation (%sO2) findings for TBI (A-C) and SAH (D-F). In (A), there are hyperintensities in all three of T2, T2*-weighted and Trace images at the impact site, with reduced CBF and %sO2 at the same region. The animal in (B) has the hemorrhagic core surrounded by vasogenic edema deep inside the brain on the right side. CBF is similarly low in the region while %sO2 recovers from the bilateral low levels to almost normal state 3h post injury. One animal had two contusion sites (C) where the CBF map and oxygenation also show reductions in the same location. While all SAH cases indicate the hypointense T2*-weighted region indicating hemorrhage and cytotoxic edema in the form of hyperintense T2 and hypointense Trace, there is a heterogeneity in the location of SAH pathology ranging from mid-brain (D), cortex (E), and base of the brain (F). The hypointense T2*-weighted region in (D) was in slices further posterior in the brain. The CBF changes are congruent with oxygenation but not as focal as the edema and hemorrhage seen in T2, T2*-weighted and Trace maps.

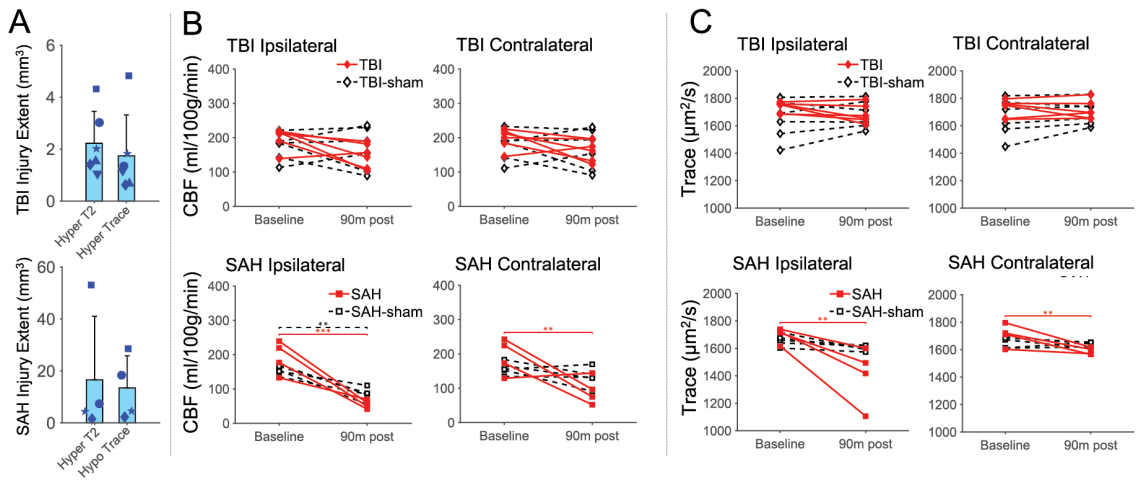

**Figure 7.** A) Extents of abnormal T2 and Trace regions in TBI and SAH injuries. Focal hyperintense T2 and Trace was the predominant observation around TBI contusions while extended hyperintense T2 and hypointense Trace sites were seen in SAH cases. Trends of B) cerebral blood flow averaged across three 1-mm ASL slices, and C) diffusion tensor Trace averaged over the brain on ipsilateral and contralateral sides (top 2-mm part of the brain, * p < 0.05, ** p < 0.01, *** p < 0.001).

# Supplementary Information

## Ultrasound and Photoacoustic Imaging

Each set of US/PA scans took approximately 20 minutes to complete and included 3D color Doppler images (with a pulse rate frequency of 3 kHz to capture blood flow velocity in the major vessels of the mouse brain), 3D maps of %sO2 obtained by unmixing the PA signal from 750 and 850 nm wavelengths, as well as underlaid 3D ultrasound B-mode images. To observe vitals such as temperature, respiration, and heart rate, we placed the mice on a ventrally heated imaging platform in the prone position and continuously monitored physiological metrics using the Vevo animal monitoring system. For these 3D acquisitions, axial slices were collected automatically by the movement of a linear actuator at 0.15 mm intervals across the entire brain.

## MRI

The following is a detailed description of magnetic resonance imaging (MRI) scans used in our study. We used a multi-echo RARE sequence for T2-mapping (TE/TR=7-77 ms/5000 ms; 200x200x400 micron, NEX=2, reps=6) and a multi-gradient echo sequence to collect T2*-weighted images (TE/TR=3.5-26 ms /55 ms; 200x200x300 micron, NEX=1, reps=6). Time of flight (TOF) angiography images were acquired using a 3D gradient echo pulse sequence (TE/TR=2.68 ms/25 ms; flip angle=20; 150 μm isotropic, NEX=1, reps=1). Our diffusion tensor imaging (DTI) approach used a single shot echo planar imaging (EPI) pulse sequence (TE/TR=41 ms/5000 ms; 250x250x400 micron; NEX=1; reps=2 with opposite phase encode direction for later blip-up-blip-down correction of geometric distortions) and b=800-1600 s/mm2 with 30 non-collinear diffusion encoding directions per shell. Finally, CBF maps were collected using a pseudo-continuous arterial spin labelling sequence (ASL) using 2D 1-shot EPI (1) (TE/TR=11 ms/4000 ms; 0.281x0.281x1.0 mm; NEX=1, reps=60; 30 labeled and 30 unlabeled). Three separate acquisitions of ASL were collected with 3 consistently placed axial slices with respect to vasculature landmarks found in TOF images: posterior cerebral artery and Circle of Willis, bregma (anterior closing of Circle of Willis), and 1 mm anterior from bregma.

## Imaging Analysis and Post Processing

### 2D Slice Comparison

To compare images from MRI and US/PA we used similar processing methods to avoid alternative interpretations between the two modalities. For co-localization between MRI and US/PA scans, we extracted axial US/PA slices associated with bregma and PCAs based on the anatomical B-mode and color Doppler images. The corresponding ASL slice locations were used as a reference to select co-localized images from the other MRI sequences. Specifically, we used the spatial coordinates from the raw MRI data from the ASL scans, to extract the corresponding T2, T2*-weighted, and DTI Trace slices at the same locations assuming no significant motion, which we confirmed by inspection.

**US/PA**

We processed 3D %sO2 maps and color Doppler datasets using the accompanying B-mode images as anatomical reference. For %sO2 maps, we drew 3D regions of interest (ROI) over the area covering approximately 2 mm under the skull for whole brain, left hemisphere, and right hemisphere using the VevoLab software's 3D ROI tool. To quantify %sO2 and limit the quantification to voxels with high signal to noise ratio, we selected a total hemoglobin threshold of 40%. We exported the data, containing single slice and total volume averages of %sO2 in the ROI, to MATLAB (Mathworks, MA) for analysis. We computed %sO2 values in 10.5 mm (anterior-posterior) of brain tissue starting from 3 mm anterior to bregma. We excluded the olfactory bulb and the 2-3 mm of posterior brain from the volumetric analysis due to loss of signal to noise ratio in the photoacoustic scans. We weighted %sO2 values by the area of each axial region of interest (ROI) over the two sides of the brain – ipsilateral and contralateral to the side of injury.

We then analyzed the color Doppler images offline using a custom script in MATLAB. To that end, we exported raw color Doppler and B-mode images from the US/PA instrument and aligned them based on geometrical offsets reported in their corresponding header file. We then created 3D brain masks by drawing manual ROIs over the brain in 6-8 axial slices spaced out over the brain in anterior-posterior direction and swept splines over them in orthogonal planes. In addition, we manually selected three voxels in the midsagittal plane (one near the olfactory bulb and two in the posterior region) for each scan and used them to divide the brain masks into two hemispheres. We applied the masks to the color Doppler images to obtain blood flow velocities in the ipsilateral and contralateral sides of injury.

Finally, we multiplied the absolute value of each voxel's measured velocity by voxel volume and summed over the ipsilateral and contralateral ROIs to obtain the velocity-weighted volume of flow (VVF) metric.

**MRI**

For MR image processing and quantitative mapping, we used custom software tools written in MATLAB: T2*-weighted images were the average of all echoes in the multi-echo multi-gradient echo sequence. T2 maps were the result of exponential fitting of all echoes, and cerebral blood flow (CBF, mL/100g/min) maps were calculated from ASL data using an algorithm outlined in (1) with a fixed inversion efficiency of 70%. We used TORTOISE for image corrections and diffusion tensor fitting of DTI data (2) and extracted the Trace of the diffusion tensor (µm2/s) as a measure of diffusivity.

Region of interest masks were drawn semi-manually to include the 2-3 mm cortical region drawn in the %sO2 analysis for Trace and CBF images. To that end, we created a midplane for each animal specimen by selecting an anterior voxel, and two posterior voxels at the dorsal and ventral sides using ITK-SNAP software (3). Using a MATLAB script, we used these three voxels to create a mid-sagittal plane for each scan, and divided the image based on this calculated midplane to create left and right masks. We developed and used a custom script that included only the 2 mm dorsal part of the brain at each axial Trace and CBF slice and averaged Trace and CBF over ipsilateral and contralateral sides, similar to the %sO2 ROIs. This approach also ensured high signal to noise ratio from locations closest to the receiver coil and US transducer for MRI and US/PA, respectively. For Trace values, we only included values lower than 2000 µm2/s in this analysis to avoid reading confounding values from cerebrospinal fluid and white matter.

We also created 3D ROIs for pathology in T2 and Trace maps for SAH and TBI using ITK-SNAP by locating abnormal signal within the brains. We were not blinded to the study groups but did confirm ROIs by at least two people. When no visible injury could be detected, we merely reported the case but did not draw a corresponding 3D ROI. To quantify the extent of local injuries, we counted the number of voxels that made up the 3D ROI for each injury and multiplied that by the voxel volume.

We also used publicly available MATLAB codes (4,5) for better visualization of results.

Table S1: Statistical summary of between-group analysis after repeated-measure analysis of variance (ANOVA)

|  | Grouping | F | pValue | $\eta^2$ |
|---|---|---|---|---|
| Oxygenation (PA) | TBI/TBI-sham: ipsilateral | 40.23 | **<0.001** | **0.8** |
|  | TBI/TBI-sham: contralateral | 16.12 | **<0.001** | **0.62** |
|  | SAH/SAH-sham: ipsilateral | 11.29 | **0.007** | **0.62** |
|  | SAH/SAH-sham: contralateral | 3.09 | 0.115 | - |
| VVF (CD) | TBI/TBI-sham: ipsilateral | 6.71 | **0.013** | **0.4** |
|  | TBI/TBI-sham: contralateral | 3.89 | **0.04** | **0.28** |
|  | SAH/SAH-sham: ipsilateral | 3.93 | 0.055 | - |
|  | SAH/SAH-sham: contralateral | 2.13 | 0.176 | - |
| CBF (CASL) | TBI/TBI-sham: ipsilateral | 2.59 | 0.138 | - |
|  | TBI/TBI-sham: contralateral | 0.8 | 0.393 | - |
|  | SAH/SAH-sham: ipsilateral | 6.19 | **0.042** | **0.47** |
|  | SAH/SAH-sham: contralateral | 3.93 | 0.088 | - |
| Trace (DTI) | TBI/TBI-sham: ipsilateral | 5.46 | **0.042** | **0.35** |
|  | TBI/TBI-sham: contralateral | 2.35 | 0.157 | - |
|  | SAH/SAH-sham: ipsilateral | 10.17 | **0.015** | **0.59** |
|  | SAH/SAH-sham: contralateral | 8.08 | **0.025** | **0.54** |

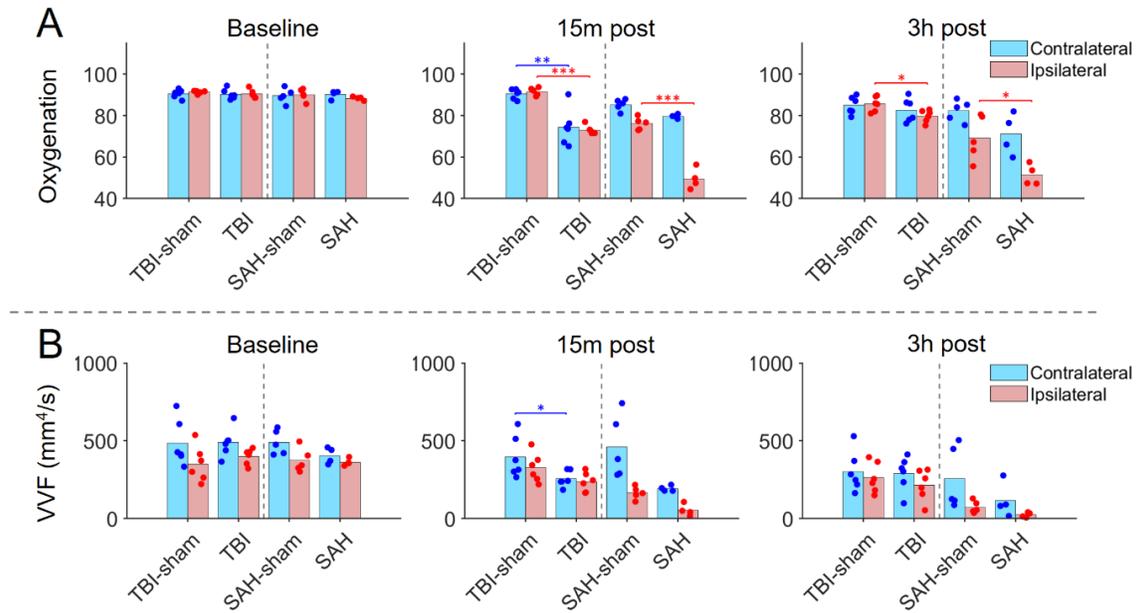

Figure S1. Injury-vs-control comparisons for A) brain blood oxygenation and B) velocity-weighted volume of flow (VVF) (* $p < .05$, ** $p < .01$, *** $p < .001$)

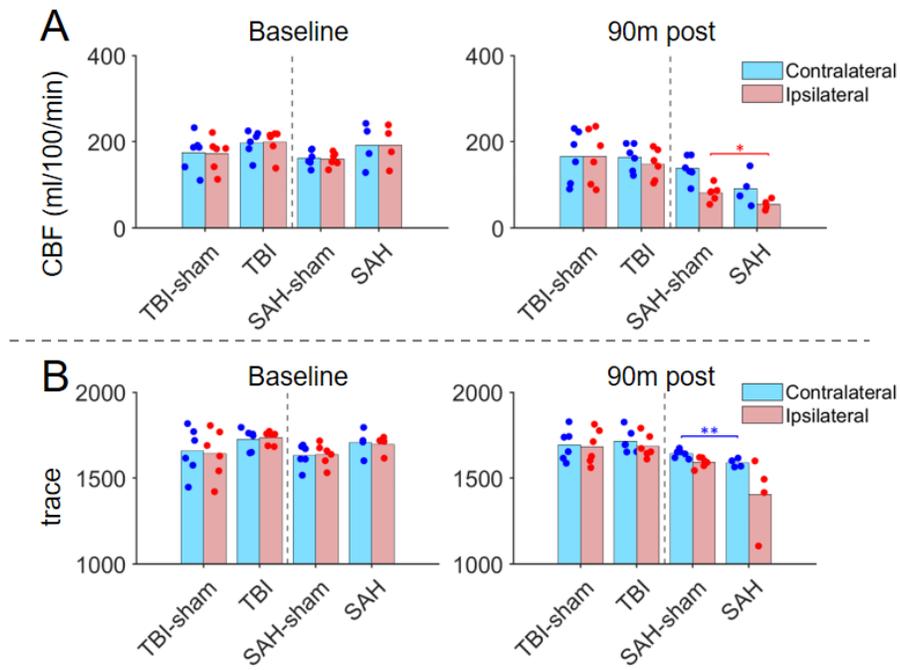

Figure S2. Injury-vs-control comparisons for A) cerebral blood flow (CBF) and B) trace of diffusion tensor (* $p < .05$, ** $p < .01$, *** $p < .001$)

# Supplementary Information References

1. Hirschler L. Developments in preclinical arterial spin labeling. Université Grenoble Alpes (ComUE); 2017.

2. Irfanoglu MO, Nayak A, Jenkins J, Pierpaoli C. TORTOISE v3: Improvements and new features of the NIH diffusion MRI processing pipeline. In: Proceedings of the 25th annual meeting of ISMRM presented at the international society for magnetic resonance in medicine. 2017.

3. Yushkevich PA, Piven J, Hazlett HC, Smith RG, Ho S, Gee JC, et al. User-guided 3D active contour segmentation of anatomical structures: significantly improved efficiency and reliability. Neuroimage. 2006;31(3):1116–28.

4. Rob Campbell (2022). raacampbell/sigstar (https://github.com/raacampbell/sigstar), GitHub. Retrieved February 24, 2022.

5. Simon Musall (2022). stdshade (https://www.mathworks.com/matlabcentral/fileexchange/29534-stdshade), MATLAB Central File Exchange. Retrieved February 24, 2022.